\documentclass[conference]{IEEEtran}
\IEEEoverridecommandlockouts

\usepackage[cmex10]{amsmath}
\usepackage{array}
\usepackage[usenames]{color}
\usepackage{epsfig}

\newtheorem{lemma}{Lemma}

\newcommand{\qed}{\hspace*{\fill} \fbox{} \par \vspace{\baselineskip}}
\newcommand{\WH}{{\rm WH}}

\DeclareMathOperator{\E}{E} 

\begin{document}

\title{Non-binary LDPC decoding using truncated messages in the Walsh-Hadamard domain}

\author{\IEEEauthorblockN{Jossy Sayir}
\IEEEauthorblockA{University of Cambridge}
\thanks{Funded in part by the European Research Council under ERC grant
agreement 259663 and by the FP7 Network of Excellence NEWCOM\#
under grant agreement 318306.}}

\maketitle

\begin{abstract}
The Extended Min-Sum (EMS) algorithm for non-binary low-density
parity-check (LDPC)
defined over an alphabet of size $q$
operates on truncated messages of length $q'$ to achieve 
a complexity of the order $q'^2$. In contrast, Walsh-Hadamard (WH) transform
based iterative decoders achieve a complexity of the order $q\log q$,
which is much larger for $q'<<q$. In this paper, we demonstrate that
considerable savings can be achieved by letting WH based decoders
operate on truncated messages as well. We concentrate on the direct
WH transform and compute the number of operations required if only 
$q'$ of the $q$ inputs are non-zero. Our paper does not cover the
inverse WH transform and hence further research is needed to construct
WH based decoders that can compete with the EMS algorithm on complexity
terms.
\end{abstract}

\section{Introduction}
\label{sec:intro}

Extended Min-Sum (EMS) type decoders \cite{declercq2007} for non-binary LDPC codes
achieve a considerable complexity reduction with respect to the full
Sum-Product (SP)
decoder. EMS decoders work with truncated messages in the logarithmic
domain, while SP decoders typically work with full messages of length $q$ in the
probability domain, where $q$ is the alphabet size.
A rough side by side comparison gives
\begin{itemize}
\item a variable nodes of degree $d_v$ in the logarithmic domain performs
an addition of $d_v+1$ terms followed by $d_v$ substractions to compute
extrinsic messages;
\item the same variable node in the probability domain performs a 
multiplication of $d_v+1$ factors followed by $d_v$ divisions;
\item a check node of degree $d_c$ must perform $d_c$ cyclic 
convolutions of $d_c-1$ input vectors;
\item a cyclic convolution of two vectors of length $q$ computed
directly requires $q^2$ multiplications followed by $q$ sums
of $q$ terms;
\item in the probability domain, if the alphabet size is a power of 2, 
i.e., $q=2^m$, the cyclic convolution can be 
achieved by taking a Walsh Hadamard (WH) transform, then
performing extrinsic products as in a variable node, 
and then applying the inverse WH transform. This reduces
the complexity of the cyclic convolution from an order
of $q^2$ to an order of $q\log q$. 
\end{itemize}
Fitting these elements together, we see that if the EMS works
with truncated messages of length $q'<q$, then it will achieve
a lower complexity order only if $q'^2<q\log q$.

Complexity order is only relevant in the asymptotic regime.
For non-binary LDPC decoding, the asymptotic complexity for
large $q$ and $q'$ is irrelevant. The cases of interest for all 
practical purposes are for $q$ between 4 and 256. Therefore,
in order to assess the benefits of EMS decoders, it is crucial 
to get an exact comparison between the number of operations
for specific values of interest for $q$ and $q'$.

In the literature on EMS decoders \cite{declercq2007,declercq2010,declercq2013},
it is naturally
assumed that the WH based decoders in the probability domain
can only be applied to full non-truncated messages and therefore
there can be no reduction from the complexity order of $q\log q$.
In this contribution, we show that this is not exactly true
and that, while the WH transform must be applied to a full 
length message, some complexity savings can be achieved if
the full length message has been converted from a truncated
message. We proceed to define a framework for counting the
exact number of additions and minus operations that are 
required by a WH transform 
in a decoder
working with truncated messages. This paves the way for a fair
complexity comparison of the EMS with essentially equivalent WH-based
approaches.

\section{Truncated messages, logarithms, and probability distributions}
\label{sec:truncmsg}

If we work with truncated messages, it is necessary to specify
which symbols in GF($q$) the message entries correspond
to. This is not necessary for full-length messages because
the natural ordering of message entries to symbols can be assumed.
Reduced complexity approaches based on the cyclic 
convolution of truncated messages need to carry the assignment
of message entries to symbols through the operations, performing
sums in GF($q$) in parallel to the message value calculations
in order to work out the assignment of message entries to symbols
in the resulting message. Note that when applying a cyclic
convolution to two truncated messages of length $q'<q$, the 
result is likely to have more than $q'$ entries. Therefore,
selecting which of the $q'$ entries to retain in the resulting
truncated message is a non-trivial operation that is 
part of the design process for reduced complexity algorithms.

Working in the logarithmic domain with full messages is 
fully equivalent to the probability domain. If we denote
messages in the probability domain as $m_p = (p_1,p_2,\ldots,p_q)$
then the equivalent message in the logarithmic domain 
$m_l = (\lambda_1,\lambda_2,\ldots,\lambda_q)$ is defined as
\[
\lambda_i = \log p_i + \lambda_0 \text{ for $i=1\ldots q$}
\]
where $\lambda_0$ is an arbitrary constant, typically
$\lambda_0=-\log p_1$ or $\lambda_0=-\min_i\log p_i$, where
the latter ensures that all message entries are positive.
Even if the constant $\lambda_0$ is unknown, the 
probability vector can be recovered from the vector of
logarithms by normalizing so that its entries sum to 1.

For truncated messages, working in the logarithmic
domain adds a dimension of subtelty.
In the probability domain, the values of the truncated
message contain an implicit statement about the values
of the probabilities in the part of the message
that is missing. Since the probabilities over the complete
symbol alphabet must sum to 1, we know that the sum of
the probabilities in the missing part is the difference
between the sum of message values and 1, i.e., if $T$ is
the set of symbols corresponding to entries in the 
truncated message,
\[
\sum_{i\notin T} p_{i+1} = 1 - \sum_{i\in T} p_{i+1}.
\]
It is common to assume that the probabilities in the missing
part of the message are uniformly distributed, i.e.,
for $j\notin T$,
\begin{equation}
p_{j+1} = -\frac{1}{q'}\left(1-\sum_{i\in T_i}p_{i+1}\right)
\label{eq:tailprobs}
\end{equation}
where $q' = q-|T|$.
When working in the logarithmic domain, there is now
an extra degree of freedom, as the sum of probabilities
in the message is not expected to be 1, and therefore
the $\lambda_0$ cannot be recovered from the message.
If the $\lambda_0$ used in the conversion is unspecified,
the logarithmic message becomes disconnected from any
specific probability vector.
In practice, reduced complexity methods operating cyclic
convolutions on truncated messages do not bother to 
specify $\lambda_0$ and appear not to suffer from the
resulting disconnection.

Reduced complexity methods operating on truncated messages
aim to retain the symbols with highest probabilities within
their truncated message, assuming all others to be uniformly
distributed. Since the transformation into the logarithmic
domain is monotone irrespective of $\lambda_0$, retaining
the symbols with maximal $\lambda_i$ is equivalent to 
maximizing the corresponding probabilities.

\section{Reduced Complexity Walsh-Hadamard Transform}
\label{sec:rcwh}

We have seen that conceptually, methods operating on truncated
messages are assuming probability distributions with uniform 
tails on the complete symbol alphabet, where the uniform 
tail contains the symbols that are missing in the truncated message.
If we now consider the Walsh-Hadamard approach to
the cyclic convolution, we are constrained by the
fact that the WH transform cannot be applied to a truncated
message. This is because the rule that multiplication in the
WH domain is equivalent to a convolution in the time domain
only applies if the WH transform is taken over the full symbol
alphabet size. We can however replace a true complete probability
distribution by a distribution 
with a uniform tail, following the same concept as truncated
message decoders. Indeed, we can transmit truncated messages
along the edges of our decoder graph, and add a uniform tail
before the message enters the WH transform. Similarly, we
can truncate the message coming out of the WH transform
before the check node outputs it to an edge in the graph.

Therefore, any complexity reduction for WH-based decoders
operating on messages with uniform tails must answer
the following questions:
\begin{itemize}
\item Can the complexity of the WH transform be reduced
below $O(q\log q)$ when the input vector has a uniform tail?
\item Can the complexity of the {\em inverse} WH transform be
reduced below $O(q\log q)$ if we are ultimately only
interested in the largest $q'$ outputs of the transform?
\end{itemize}

We will address these questions in the following sub-sections.

\subsection{Direct WH Transform}

Let $T$ be the set of $q'$ symbol indices in a truncated message received
from the graph. We complete the message as a full-length message
$m=(p_1,\ldots,p_q)$ such that $p_i$ is the entry in the truncated
message for $i\in T$, and $p_i = p_0$ where $p_0$ is defined as in
(\ref{eq:tailprobs}). 
Let us now re-write the message as a sum
\[
m = m^{(1)} + m^{(2)}
\]
where $m^{(1)}$ is a uniform message of length $q$ with entries
\[
m^{(1)}_i = p_0 \text{ for $i=1,2,\ldots,q$}
\]
and $m^{(2)}$ is defined as
\[
\left\{\begin{array}{ll}
m^{(2)}_i = p_i-p_0 & \text{ for $i\in T$} \\
m^{(2)}_i = 0 & \text{ for $i\notin T$}.
\end{array}
\right.
\]
Since the WH transform is linear, the transform of $m$
is equal to the sum of the transforms of $m^{(1)}$ and 
$m^{(2)}$. The WH transform of the uniform 
vector $m^{(1)}$ has a nonzero component $qp_0$ in position 1
and all zeros elsewhere.
Our problem then is to estimate the complexity of 
the WH transform applied to a vector $m^{(2)}$ of length $q$
with only $q'<q$ non-zero entries. 

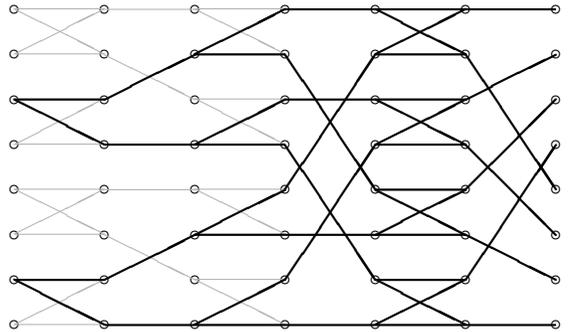
\begin{figure}[h]
\centering
\setlength{\unitlength}{.6mm}
\begin{picture}(120,70)
\newsavebox{\eightnodes}
\savebox{\eightnodes}(0,70)[lb]{
\multiput(0,0)(0,10){8}{\circle{2}}
}
\multiput(2,0)(20,0){7}{\usebox{\eightnodes}}
\definecolor{LightGray}{gray}{.7}
\color{LightGray}
\thinlines
\newsavebox{\onebutterfly}
\savebox{\onebutterfly}(20,10)[lb]{
\multiput(0,0)(0,10){2}{\line(1,0){20}}
\put(0,0){\line(2,1){20}}
\put(0,10){\line(2,-1){20}}
}
\newsavebox{\fourbutterflies}
\savebox{\fourbutterflies}(20,80)[lb]{
\multiput(0,0)(0,20){4}{\usebox{\onebutterfly}}
}
\multiput(0,0)(40,0){3}{\usebox{\fourbutterflies}}
\multiput(20,0)(0,40){2}{\line(1,0){20}}
\multiput(20,30)(0,40){2}{\line(1,0){20}}
\multiput(20,10)(0,40){2}{\line(2,1){20}}
\multiput(20,20)(0,40){2}{\line(2,-1){20}}
\multiput(60,0)(0,20){2}{\line(1,0){20}}
\multiput(60,50)(0,20){2}{\line(1,0){20}}
\multiput(60,10)(0,20){2}{\line(2,3){20}}
\multiput(60,40)(0,20){2}{\line(2,-3){20}}
\put(100,0){\line(1,0){20}}
\put(100,10){\line(2,3){20}}
\put(100,20){\line(2,-1){20}}
\put(100,30){\line(1,1){20}}
\put(100,40){\line(1,-1){20}}
\put(100,50){\line(2,1){20}}
\put(100,60){\line(2,-3){20}}
\put(100,70){\line(1,0){20}}
\color{black}
\thicklines
\put(0,10){\line(1,0){20}} \put(20,10){\line(2,1){20}}
\put(0,10){\line(2,-1){20}} \put(20,0){\line(1,0){20}}
\put(40,0){\line(1,0){20}} \put(60,0){\line(1,0){20}}
\put(40,0){\line(2,1){20}} \put(60,10){\line(2,3){20}}
\put(40,20){\line(1,0){20}} \put(60,20){\line(1,0){20}}
\put(40,20){\line(2,1){20}} \put(60,30){\line(2,3){20}}
\put(80,0){\line(1,0){20}} \put(100,0){\line(1,0){20}}
\put(80,0){\line(2,1){20}} \put(100,10){\line(2,3){20}}
\put(80,20){\line(1,0){20}} \put(100,20){\line(2,-1){20}}
\put(80,20){\line(2,1){20}} \put(100,30){\line(1,1){20}}
\put(80,40){\line(1,0){20}} \put(100,40){\line(1,-1){20}}
\put(80,40){\line(2,1){20}} \put(100,50){\line(2,1){20}}
\put(80,60){\line(1,0){20}} \put(100,60){\line(2,-3){20}}
\put(80,60){\line(2,1){20}} \put(100,70){\line(1,0){20}}
\put(0,50){\line(2,-1){20}} \put(20,40){\line(1,0){20}}
\put(0,50){\line(1,0){20}} \put(20,50){\line(2,1){20}}
\put(40,40){\line(1,0){20}} \put(60,40){\line(2,-3){20}}
\put(40,40){\line(2,1){20}} \put(60,50){\line(1,0){20}}
\put(40,60){\line(1,0){20}} \put(60,60){\line(2,-3){20}}
\put(40,60){\line(2,1){20}} \put(60,70){\line(1,0){20}}
\put(80,10){\line(2,-1){20}} %
\put(80,10){\line(1,0){20}} %
\put(80,30){\line(2,-1){20}} %
\put(80,30){\line(1,0){20}} %
\put(80,50){\line(2,-1){20}} %
\put(80,50){\line(1,0){20}} %
\put(80,70){\line(2,-1){20}} %
\put(80,70){\line(1,0){20}} %
\end{picture}
\caption{WH transform for an input vector of length 8 with only 2 non-zero elements at positions 2 and 6}
\label{fig:whdir}
\end{figure}
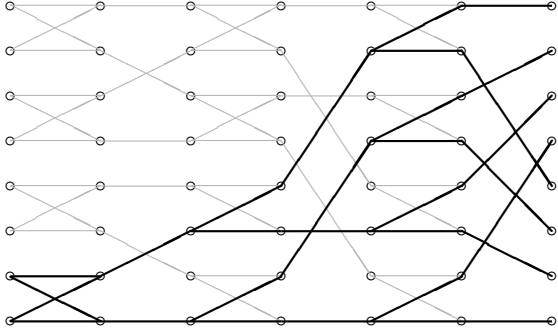
\begin{figure}[h]
\setlength{\unitlength}{.6mm}
\centering
\begin{picture}(120,70)
\savebox{\eightnodes}(0,70)[lb]{
\multiput(0,0)(0,10){8}{\circle{2}}
}
\multiput(2,0)(20,0){7}{\usebox{\eightnodes}}
\definecolor{LightGray}{gray}{.7}
\color{LightGray}
\thinlines
\savebox{\onebutterfly}(20,10)[lb]{
\multiput(0,0)(0,10){2}{\line(1,0){20}}
\put(0,0){\line(2,1){20}}
\put(0,10){\line(2,-1){20}}
}
\savebox{\fourbutterflies}(20,80)[lb]{
\multiput(0,0)(0,20){4}{\usebox{\onebutterfly}}
}
\multiput(0,0)(40,0){3}{\usebox{\fourbutterflies}}
\multiput(20,0)(0,40){2}{\line(1,0){20}}
\multiput(20,30)(0,40){2}{\line(1,0){20}}
\multiput(20,10)(0,40){2}{\line(2,1){20}}
\multiput(20,20)(0,40){2}{\line(2,-1){20}}
\multiput(60,0)(0,20){2}{\line(1,0){20}}
\multiput(60,50)(0,20){2}{\line(1,0){20}}
\multiput(60,10)(0,20){2}{\line(2,3){20}}
\multiput(60,40)(0,20){2}{\line(2,-3){20}}
\put(100,0){\line(1,0){20}}
\put(100,10){\line(2,3){20}}
\put(100,20){\line(2,-1){20}}
\put(100,30){\line(1,1){20}}
\put(100,40){\line(1,-1){20}}
\put(100,50){\line(2,1){20}}
\put(100,60){\line(2,-3){20}}
\put(100,70){\line(1,0){20}}
\color{black}
\thicklines
\put(0,10){\line(1,0){20}} \put(20,10){\line(2,1){20}}
\put(0,10){\line(2,-1){20}} \put(20,0){\line(1,0){20}}
\put(0,0){\line(2,1){20}} %
\put(0,0){\line(1,0){20}} %
\put(40,0){\line(1,0){20}} \put(60,0){\line(1,0){20}}
\put(40,0){\line(2,1){20}} \put(60,10){\line(2,3){20}}
\put(40,20){\line(1,0){20}} \put(60,20){\line(1,0){20}}
\put(40,20){\line(2,1){20}} \put(60,30){\line(2,3){20}}
\put(80,0){\line(1,0){20}} \put(100,0){\line(1,0){20}}
\put(80,0){\line(2,1){20}} \put(100,10){\line(2,3){20}}
\put(80,20){\line(1,0){20}} \put(100,20){\line(2,-1){20}}
\put(80,20){\line(2,1){20}} \put(100,30){\line(1,1){20}}
\put(80,40){\line(1,0){20}} \put(100,40){\line(1,-1){20}}
\put(80,40){\line(2,1){20}} \put(100,50){\line(2,1){20}}
\put(80,60){\line(1,0){20}} \put(100,60){\line(2,-3){20}}
\put(80,60){\line(2,1){20}} \put(100,70){\line(1,0){20}}
\end{picture}
\caption{WH transform for an input vector of length 8 with only 2 non-zero elements at positions 1 and 2}
\label{fig:whdir2}
\end{figure}

An example of this is illustrated in Figure~\ref{fig:whdir},
where the WH transform is applied to a vector of length 8 with
only 2 non-zero elements. The bold lines in the graph correspond
to edges transporting non-zero elements, while the gray edges
transport only zeros. Instead of the usual $3\times 8$
additions and $3\times 4$ minus operations
required by the WH transform, we see that only 8 additions
and 10 minus operations are performed in this case.
A WH butterfly processing two zeros does not need to be 
activated at all. A WH butterfly processing one non-zero 
element and a zero requires only a copy and possibly a minus
operation but no addition. Only WH butterflies receiving two
non-zero elements perform two additions and one minus each.
The number of additions and minus operations can vary, as illustrated in
Figure~\ref{fig:whdir2}, where only 2 additions and 1 minus
operation are required for a different configuration of
2 non-zero elements in a length 8 input vector.

As these figures demonstrate, the number of operations is
a random variable that depends on the position of the non-zero
elements in the input vector. Let us denote as $Z = Z_1Z_2\ldots Z_q$
a vector of indicator random variables that are 1 if 
the corresponding entry is non-zero and 0 if the corresponding
entry is 0. The number of non-zero elements $q'$ in an input
vector is the Hamming weight $w(Z)$ of the corresponding vector
of indicator variables. We will assume that all
patterns of non-zero input elements are equally probable, e.g.,
$P_Z(z)= 2^{-q}$ for any $z$. 
The number of additions $A$ and the number of minus operations $M$
depend only on the vector
of indicator random variables, i.e., $A = f_A(Z)$ and $M=f_M(Z)$. We are
interested in evaluating the expectated number of
operations $\E[A|w(Z)=q']$ and $E[M|w(Z)=q']$ in the WH transform.

We follow two approaches, one approximate and the other exact.
The approximate approach assumes that $Z$ is the output
of a Bernoulli process with parameter $p=q'/q$. This
means that the number of non-zero entries is now a random
variable rather than being fixed, but its expected value is
equal to $q'$, and all sequences of weight $q'$ remain 
equi-probable, even though we are assuming the wrong sequence
probabilities. If we consider the first
layer of butteflies in the WH transform, we note that both outputs of a
butterfly will be non-zero if any or both of its inputs
are non-zero\footnote{Throughout this section, we
neglect the possibility that two inputs to an adder
would be non-zero and equal with opposite signs, in which 
case one of the outputs of the butterfly could in theory be zero.
Since these are real numbers we will assume that the
probability of this event is zero.\label{footnote:nonzero}}. Therefore,
we can write a recursive formula for the probability
of a non-zero entry at the output of a layer of butterflies
given its input probability
\begin{equation}
p_{i+1} = 1 - (1-p_i)^2 \text{ for $i=1,2,\ldots$}
\label{eq:recursive_probs}
\end{equation}
where we note $p_1 = p$ for the input probability of the
first layer. Of course the outputs of a layer are not
Bernoulli, since non-zero outputs always come in pairs.
We will make a further approximation in assuming that the
interleavers between layers of butterflies are random, resulting in
Bernoulli inputs to each layer, so that we can safely apply
(\ref{eq:recursive_probs}) to all layers in the WH transform.
A butterfly with two non-zero entries will perform two 
additions and one minus operation; a butterfly with
one non-zero entry will perform no additions and possibly one
minus; and a butterfly with zero entries performs no
operations at all. Therefore, we can express the expected number
of additions for a layer as
\begin{equation}
\E[A_i] \approx \frac{q}{2}(2p_i^2) = qp_i^2
\label{eq:adds_layer}
\end{equation}
where $2p_i^2$ is the expected number of additions per
butterfly and $q/2$ is the number of butterflies per layer.
Similarly, we can express the expected number of minus
operations for a layer as
\begin{equation}
\E[M_i] \approx \frac{q}{2}[p_i^2+p_i(1-p_i)] = qp_i/2.
\label{eq:minuses_layer}
\end{equation}
By applying (\ref{eq:recursive_probs}) recursively
and (\ref{eq:adds_layer}) and (\ref{eq:minuses_layer})
to each layer, we can calculate a simple approximation
to the number of additions and minus operations required
by the length $q$ WH transform with $pq$ non-zero entries.
These figures will not be exact because they rely on two
approximations, namely replacing the exact number of 
non-zero entries by an expectation, and assuming that 
the interleavers between layers in the WH transform
are random. Table~\ref{table:nops_wh64} at the end of this section shows the
results obtained with the approximate method versus the
exact values for $q=64$ . The approximations appear to be slightly
lower than the exact values. The approximate approach has
the advantage that it is much easier to evaluate and does
not require to evaluate any factorials.

\begin{table}[h]
\caption{Excact numbers of operations for $q'$ non-zero inputs in a
length $q=2$ WH transform, and exact number of additions and minus operations in a length $q=4$
WH transform}
\label{table:nop_exact}
\centering
\begin{tabular}{|c||c|c|}
\hline
$q'$ & $\E[A]$ & $\E[M]$ \\ \hline \hline
0 & 0 & 0 \\ \hline
1 & 0 & $1/2$ \\ \hline
2 & 2 & 1 \\ \hline
\end{tabular}
\medskip

\begin{tabular}{|c|c|c||c|c|c|c||c|}
\hline
$q'$ & $q'_L$ & $q'_R$ & $\frac{P(q_L,q_R)}{P(q')}$ & $\E[A_L]$ & $\E[A_R]$ & $+q$? & $E[A]$ \\ \hline \hline
0 & 0 & 0 & 1  & 0 & 0 & 0 & 0 \\ \hline
1 & 1 & 0 & $1/2$  & 0 & 0 & 0 & \\
  & 0 & 1 & $1/2$  & 0 & 0 & 0 & 0 \\ \hline
2 & 2 & 0 & $1/6$  & 2 & 0 & 0 & \\
  & 1 & 1 & $2/3$  & 0 & 0 & 4 & \\
  & 0 & 2 & $1/6$  & 0 & 2 & 0 & $10/3$ \\ \hline
3 & 2 & 1 & $1/2$  & 2 & 0 & 4 & \\
  & 1 & 2 & $1/2$  & 0 & 2 & 4 & 6 \\ \hline
4 & 2 & 2 & 1  & 2 & 2 & 4 & 8 \\ \hline
\end{tabular}
\medskip

\begin{tabular}{|c|c|c||c|c|c|c||c|}
\hline
$q'$ & $q'_L$ & $q'_R$ & $\frac{P(q_L,q_R)}{P(q')}$ & $\E[M_L]$ & $\E[M_R]$ & $+q/2$? & $E[M]$ \\ \hline \hline
0 & 0 & 0 & 1  & 0 & 0 & 0 & 0 \\ \hline
1 & 1 & 0 & $1/2$  & $1/2$ & 0 & 0 & \\
  & 0 & 1 & $1/2$  & 0 & $1/2$ & 2 & $3/2$ \\ \hline
2 & 2 & 0 & $1/6$  & 1 & 0 & 0 & \\
  & 1 & 1 & $2/3$  & $1/2$ & $1/2$ & 2 & \\
  & 0 & 2 & $1/6$  & 0 & 1 & 2 & 3 \\ \hline
3 & 2 & 1 & $1/2$  & 1 & $1/2$ & 2 & \\
  & 1 & 2 & $1/2$  & $1/2$ & 1 & 2 & $7/2$ \\ \hline
4 & 2 & 2 & 1  & 1 & 1 & 2 & 4 \\ \hline
\end{tabular}
\vspace{-.6cm}
\end{table}
To compute the exact number of operations, 
we will make use of the following two lemmas:
\begin{lemma}
Let $\WH_i(X)$ denote the $i$-th element of the Walsh-Hadmard
transform of the vector $X$. Let $X_i^j$ denote the portion of $X$
starting at index $i$ and ending at index $j$. We have
\vspace{-.1cm}
\begin{equation}
\WH_i(X) = \begin{cases}
\WH_i(X_1^{q/2}) + \WH_i(X_{q/2+1}^q) \\
\;\; \text{if $i=1\ldots q/2$} \\
\WH_{i-q/2}(X_1^{q/2}) - \WH_{i-q/2}(X_{q/2}^{q})  \\
\;\; \text{if $i=q/2+1\ldots q$}
\end{cases}
\label{eq:fht}
\end{equation}
\label{lemma:fht}
\end{lemma}
{\em Proof:} this follows directly from he definition of the WH
matrix as a successive Kronecker product of the matrix
\[
W_2 = \left[\begin{array}{cc}1&1\\ 1&-1\end{array}\right]
\]
with itself. Decomposing the WH transform of length $q$ as 
the Kronecker product of $W_2$ with the WH transform of
length $q/2$ gives the expression in (\ref{eq:fht}).
\qed
Lemma~\ref{lemma:fht} essentially provides the basis for the 
``fast Hadamard transform'' (FHT) that gives us a complexity
of $q\log q$ for what would otherwise be a matrix multiplication
with a Hadamard matrix, which has a complexity of $q^2$.
\begin{lemma}
Let the random variable $A$ be the number of additions required in a FHT,
$M$ the number of minus operations, $Q$ the length of the
transform and $Q'$ the number of non-zero elements, then
\begin{align}
&\E[A|Q=q,Q'=q'] = \\ \nonumber
&\;\; \sum_{\stackrel{q_L=\max\{0,q'-q/2\}}{q_R=q'-q_L}}^{q_L=\min\{q',q/2\}}
\frac{\binom{q/2}{q_L}\binom{q/2}{q_R}}{\binom{q}{q'}}
\left(\vphantom{\frac{1}{2}}
\E\left[A|Q=q/2,Q'=q_L\right] \right. \\ \nonumber
&\;\;\left. +\E\left[A|Q=q/2,Q'=q_R\right]+q\left(1-\delta(q_Lq_R)\right)
\vphantom{\frac{1}{2}}\right) 
\end{align}
\begin{align}
&E[M|Q=q,Q'=q'] = \\ \nonumber
&\;\; \sum_{\stackrel{q_L=\max\{0,q'-q/2\}}{q_R=q'-q_L}}^{q_L=\min\{q',q/2\}}
\frac{\binom{q/2}{q_L}\binom{q/2}{q_R}}{\binom{q}{q'}}
\left(\vphantom{\frac{1}{2}}
\E\left[M|Q=q/2,Q'=q_L\right] \right. \\ \nonumber
&\;\; \left. +\E\left[M|Q=q/2,Q'=q_R\right]+\frac{q}{2}\left(1-\delta(q_L)\right)
\vphantom{\frac{1}{2}}\right)
\end{align}
where $\delta(.)$ denotes the Kronecker delta function whose value is 1 when
its input is 0 and 0 otherwise.
\label{lemma:additions}
\end{lemma}
{\em Proof:}
Let us split the variable $Q'$ into two random variables, $Q_L$ for 
the number of non-zero elements in the left half of the input
vector and $Q_R$ for the number of non-zero elements in the right half
of the input vector. Obviously, $Q'=Q_L+Q_R$. Furthermore,
\begin{align*}
&\E[A|Q=q,Q'=q'] =\\
&\;\; \sum_{q_L}\sum_{q_R}\E[A|Q,Q_L,Q_R,Q'=q,q_L,q_R,q']
P(q_L,q_R|q,q') \\
&\;\;\;=\sum_{\stackrel{q_L}{q_R=q'-q_L}}\E[A|Q=q,Q_L=q_L,Q_R=q_R]P(q_L|q,q').
\end{align*}
Let us consider the probabilities $P(q_L|q,q')$. They can only
be non-zero for consistent values of $q_L$ with respect to 
$q$ and $q'$: $q_L$ can be at most equal to $q'$ since the left half
of the input vector cannot have more non-zero elements than the whole
vector. It can also be at most $q/2$ since it cannot have more non-zero
elements than the left half has positions. Furthermore, if $q'>q/2$, 
$q_L$ must be at least equal to $q'-q/2$ in order for $q_R$ to 
remain below $q/2$. This justifies the upper and lower bound
in the summation in Lemma~\ref{lemma:additions}. We can
now write
\begin{align*}
P(q_L|q,q') &= \frac{P(q_L,q'|q)}{P(q'|q)}\\
&= \frac{P(Q_L=q_L,Q_R=q'-q_L|q)}{P(q'|q)}.
\end{align*}
Since all configurations of non-zero elements are assumed to be
equally likely, i.e., equal to $2^{-q}$, the probabilities in the
last expression can be obtained by counting the sequences fulfilling
the conditions on $Q_L$, $Q_R$ and $Q'$, so
\[
\begin{cases}
P(Q_L=q_L,Q_R=q_R|q) = \binom{q/2}{q_L}\binom{q/2}{q_R}2^{-q} \\
P(Q'=q'|q) = \binom{q}{q'}2^{-q}.
\end{cases}
\]

Now let us consider the expected value $E[A|Q=q,Q_L=q_L,Q_R=q_R]$. 
Lemma~\ref{lemma:fht} shows that every element in the WH transform
can be obtained as a sum of an element in the WH transform of the
left half with an element in the WH transform of the right half. 
Therefore, the number of additions required is the number of additions
in the WH transforms of the left and right half plus the extra
addition required to sum them. If $Q_L\neq 0$ and $Q_R\neq 0$, 
then all values in the WH transform of the halves will be non-zero
(see Footnote~\ref{footnote:nonzero}) and we will need $q$ extra
additions. Otherwise, i.e. if $Q_L=0$ or $Q_R=0$, we will need no
extra additions as one of the terms in the sum will always be zero.

\begin{figure}[h]
\centering
\includegraphics[width=0.8\columnwidth,trim=80 230 100 220]{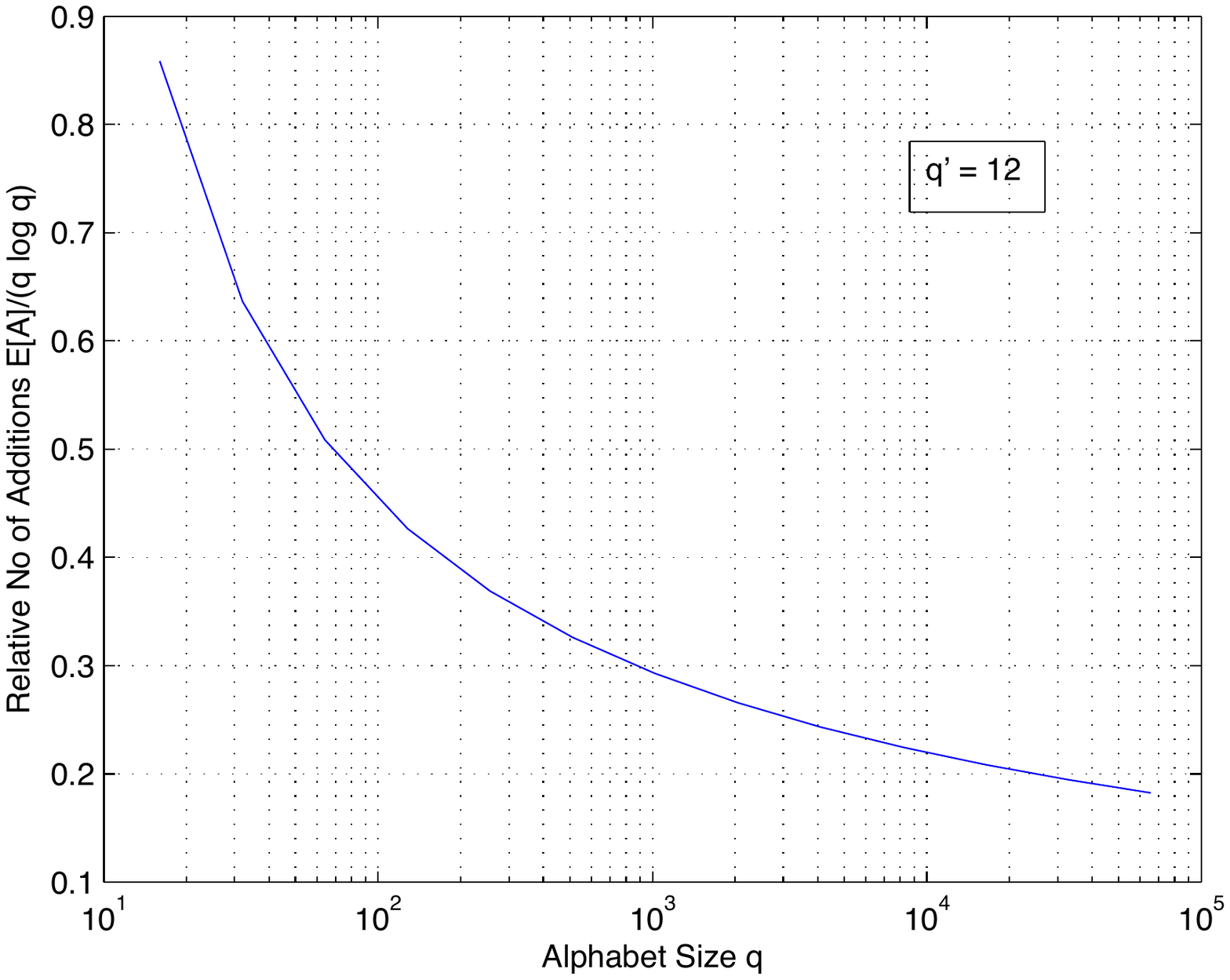}
\caption{Approximate relative expected number of additions for alphabet sizes varying from 16 to $2^{16}$ and 12 non-zero inputs, i.e., $q'=12$}
\label{fig:additions-wh}
\end{figure}
The proof of the expression for the number of minus operations
follows the same arguments, except that the number of extra
minus operations required to put the two half transforms together
is $q/2$ and is always necessary when $Q_R\neq 0$, even
if $Q_L=0$, in which case there are no sums but minus
operations are still necessary to compute $WH_i(X)$ for
$i=q/2+1,\ldots ,q$.
\qed

Lemma~\ref{lemma:additions} enables us to count operations
in a recursive manner by building up the table of operations
required for a length $q$ WH transform based on a pre-computed
number of operations required for a length $q/2$ WH transform.
Table~\ref{table:nop_exact}
illustrates this process
for the WH transform of length 4, using the number
of operations for the length 2 WH transform. 
This procedure can be extended to any length
$q$ for which we are able to compute binomial coefficients 
accurately, and can be implemented as a recursive
computer program if required. For larger values of $q$, we can 
use the approximation described previously.

\begin{table}[h]
\caption{Number of operations in the WH transform for $q=64$ and
various numbers $q'$ of non-zero entries}
\label{table:nops_wh64}
\centering
\begin{tabular}{|c||c|c||c|c|}
\hline
& \multicolumn{2}{|c|}{Approximations} & \multicolumn{2}{|c|}{Exact numbers}
\\ \hline
$q'$ & $\E[A]$ & $\E[M]$ & $\E[A]$ & $\E[M]$ \\ \hline \hline
1 & 14.4 & 27.0 & 0.0 & 31.5 \\ \hline
2 & 40.6 & 47.1 & 43.3 & 52.7 \\ \hline
3 & 67.0 & 62.5 & 74.6 & 67.8 \\ \hline
4 & 90.2 & 74.6 & 98.5 & 79.2 \\ \hline
5 & 110.2 & 84.4 & 117.7 & 88.4 \\ \hline
6 & 127.3 & 92.6 & 133.8 & 95.9 \\ \hline
7 & 142.1 & 99.5 & 147.7 & 102.4 \\ \hline
8 & 155.1 & 105.5 & 160.0 & 108.0 \\ \hline
9 & 166.6 & 110.8 & 170.9 & 113.0 \\ \hline
10 & 177.1 & 115.5 & 180.9 & 117.4 \\ \hline
11 & 186.6 & 119.8 & 190.0 & 121.5 \\ \hline
12 & 195.3 & 123.7 & 198.4 & 125.2 \\ \hline
13 & 203.4 & 127.2 & 206.2 & 128.6 \\ \hline
14 & 211.0 & 130.5 & 213.5 & 131.8 \\ \hline
15 & 218.0 & 133.5 & 220.4 & 134.7 \\ \hline
16 & 224.7 & 136.3 & 226.8 & 137.4 \\ \hline
17 & 231.0 & 139.0 & 233.0 & 140.0 \\ \hline
18 & 236.0 & 141.5 & 238.8 & 142.4 \\ \hline
19 & 242.7 & 143.8 & 244.4 & 144.7 \\ \hline
20 & 248.1 & 146.0 & 249.7 & 146.8 \\ \hline
21 & 253.3 & 148.15 & 254.8 & 148.9 \\ \hline
22 & 258.3 & 150.1 & 259.7 & 150.8 \\ \hline
23 & 263.1 & 152.0 & 264.4 & 152.7 \\ \hline
24 & 267.7 & 153.9 & 268.9 & 154.5 \\ \hline
32 & 299.7 & 165.9 & 300.5 & 166.5 \\ \hline
64 & 384 & 192 & 384 & 192 \\ \hline
\end{tabular}
\end{table}

The results in Table~\ref{table:nops_wh64} show that substantial
savings can be achieved for $q=64$ when $q'$ is smaller than about
16, with a ballpoint figure of approximately 50\% savings for additions
from $q\log q$ for $q' = 11$. The last line for $q'=64$ corresponds
to full-length messages with no non-zero entries, with exactly
$q\log q$ additions and $q\log q/2$ minus operations.
Figure~\ref{fig:additions-wh} shows the number of additions
relative to $q\log q$ for a fixed truncated size $q'=12$, 
demonstrating that the savings improve as the alphabet size grows.

\subsection{Inverse WH Transform}

A similar approach could be adopted to count the number
of operations in an inverse WH transform when only a portion of the
output message needs to be computed. However, this approach assumes
that we know which of the output symbols will be in the truncated
message and which symbols will be left out and assigned
to the uniform tail. The obvious way to select symbols for the
truncated message is to retain
the $q'$ symbols with the largest probability. However, this requires
to compute all $q$ values in order to decide
which $q'$ values are the largest. Hence, reducing the complexity
of the inverse WH transform is a harder problem, that lies
outside the scope of this paper. It may be possible to reduce the
complexity based on novel techniques
proposed in \cite{vetterli2013}.

\section{Conclusion}

We have presented evidence to the fact that using truncated
messages in WH-based iterative decoders for non-binary codes
can achieve gains with respect to full WH decoders. 
At this point, we are unable to conclude whether WH based 
decoders operating on truncated messages may compete with
EMS decoders, because constraint
nodes need to take an inverse WH transform for every outgoing
message. We have as of yet no conclusive evidence that
the complexity of the inverse WH transform can be reduced if
the target is a truncated message. Hence, currently there is no
doubt that the EMS and its many variants is the most efficient
known algorithm for decoding non-binary LDPC codes.
We believe however that WH based decoders should be investigated further
as there may be a way to achieve comparable complexity 
with methods operating in the WH domain if reduced complexity
inverse WH transforms can be devised.



%

\bibliographystyle{IEEEtran}


\begin{thebibliography}{1}
\providecommand{\url}[1]{#1}
\csname url@samestyle\endcsname
\providecommand{\newblock}{\relax}
\providecommand{\bibinfo}[2]{#2}
\providecommand{\BIBentrySTDinterwordspacing}{\spaceskip=0pt\relax}
\providecommand{\BIBentryALTinterwordstretchfactor}{4}
\providecommand{\BIBentryALTinterwordspacing}{\spaceskip=\fontdimen2\font plus
\BIBentryALTinterwordstretchfactor\fontdimen3\font minus
  \fontdimen4\font\relax}
\providecommand{\BIBforeignlanguage}[2]{{%
\expandafter\ifx\csname l@#1\endcsname\relax
\typeout{** WARNING: IEEEtran.bst: No hyphenation pattern has been}%
\typeout{** loaded for the language `#1'. Using the pattern for}%
\typeout{** the default language instead.}%
\else
\language=\csname l@#1\endcsname
\fi
#2}}
\providecommand{\BIBdecl}{\relax}
\BIBdecl

\bibitem{declercq2007}
\BIBentryALTinterwordspacing
D.~Declercq and M.~P. Fossorier, ``Decoding algorithms for nonbinary {LDPC}
  codes over {GF(q)},'' \emph{{IEEE} Trans. Commun.}, vol.~55, no.~4, pp.
  633--643, Apr. 2007. [Online]. Available:
  \url{http://publi-etis.ensea.fr/2007/DF07"}
\BIBentrySTDinterwordspacing

\bibitem{declercq2010}
A.~Voicila, D.~Declercq, F.~Verdier, M.~Fossorier, and P.~Urard,
  ``{Low-Complexity Decoding for non-binary LDPC Codes in High Order Fields},''
  \emph{IEEE Trans. on Commun.}, vol.~58, no.~5, pp. 1365--1375, May 2010.

\bibitem{declercq2013}
E.~Li, D.~Declercq, and K.~Gunnam, ``{Trellis based Extended Min-Sum Algorithm
  for Non-binary LDPC codes and its Hardware Structure},'' \emph{in IEEE Trans.
  Communications}, vol.~61, no.~7, pp. 2600--2611, July 2013.

\bibitem{vetterli2013}
\BIBentryALTinterwordspacing
R.~Scheibler, S.~Haghigahatshoar, and M.~Vetterli, ``A {F}ast {H}adamard
  {T}ransform for signals with sub-linear sparsity in the transform domain,''
  2013, arXiv pre-print. [Online]. Available:
  \url{http://arxiv.org/abs/1310.1803}
\BIBentrySTDinterwordspacing

\end{thebibliography}

\end{document}